\documentclass{article}
\usepackage{spconf,amsmath,graphicx,hyperref}
\usepackage{amssymb}
\usepackage{subfigure}
\usepackage{graphicx}

\title{A Unified SVD-Modal Solution for Sparse Sound Field Reconstruction with Hybrid Spherical-Linear Microphone Arrays}
\name{Shunxi Xu$^{1}$, Thushara Abhayapala$^{2}$, Craig T. Jin$^{1}$}

\address{
    $^{1}$Computing and Audio Research Lab, The University of Sydney, NSW, Australia\\
    $^{2}$Audio \& Acoustic Signal Processing Group, The Australian National University, ACT, Australia
}
\begin{document}
%
\maketitle
\begin{abstract}
We propose a data-driven sparse recovery framework for hybrid spherical–linear microphone arrays using singular value decomposition (SVD) of the transfer operator. The SVD yields orthogonal microphone and field modes, reducing to spherical harmonics (SH) in the SMA-only case, while incorporating LMAs introduces complementary modes beyond SH. Modal analysis reveals consistent divergence from SH across frequency, confirming the improved spatial selectivity. Experiments in reverberant conditions show reduced energy-map mismatch and angular error across frequency, distance, and source count, outperforming SMA-only and direct concatenation. The results demonstrate that SVD-modal processing provides a principled and unified treatment of hybrid arrays for robust sparse sound-field reconstruction.
\end{abstract}

\section{Introduction}
\label{sec: introduction}
Spherical microphone arrays (SMAs) provide a panoramic view of the sound field and have enabled a wide range of spatial analysis methods. Classical techniques include beamforming for energy map estimation \cite{park2005sound}, as well as high-resolution approaches such as EB-MUSIC \cite{rafaely2010spherical} and EB-ESPRIT \cite{5946342}. More recently, sparse recovery (SR) methods have gained attention as a means to enhance the spatial resolution of SMA recordings \cite{jin2020perspectives}. By modeling the sound field as a sparse superposition of a few dominant plane waves, SR achieves finer spatial detail than conventional beamforming, thereby enabling improved resolution and more accurate source localization. To further improve robustness in practical environments, several extensions have been explored, including independent component analysis (ICA) with compressive sensing (CS) fusion \cite{noohi2013direction}, subspace pre-processing \cite{epain2013super}, and spatial priming \cite{noohi2015super}.

Nevertheless, SMA-based SR methods remain constrained by the finite spherical harmonic (SH) order, which limits resolution at low frequencies and introduces aliasing at high frequencies. Linear microphone arrays (LMAs), by contrast, offer strong directional sensitivity along their array axis \cite{hu2014near}. Despite their simple and compact geometry, they provide spatial detail that complements SMA observations, making them appealing for hybrid array designs. Previous studies have shown that direct concatenation of multiple SMAs can enlarge the effective spatial aperture and improve interpolation \cite{tang2022wave} \cite{emura2017sound}. However, this strategy does not extend well to SMA–LMA combinations. Unlike SMAs, LMA channels are highly sensitive to room reflections \cite{7776154} \cite{galindo2017microphone}, and when concatenated directly, their degraded observations introduce spurious components that compromise SR performance.

To address this, our earlier work proposed a residue refinement (RR) approach, where an initial SMA estimate is refined using residual LMA information, yielding substantial performance gains \cite{Xu2025}. In this paper, we present a more principled framework that models the SMA–LMA combination as a single array and derives a unified modal solution via singular value decomposition (SVD). The resulting modal basis generalizes SH processing, reduces to SH for SMA-only setups, and provides stable, data-driven modes for sparse recovery. Modal analysis demonstrates the improved spatial resolution of the hybrid SVD basis. Results in reverberant conditions highlight the robustness of the proposed method and demonstrate the broader potential of modal analysis as a principled tool for advancing sound-field analysis with hybrid arrays.

\section{Background}
\label{sec: background}

\subsection{Sparse Plane-Wave Decomposition}
\label{subsec: sparse plane wave decomposition}

We address the problem of reconstructing a three-dimensional sound field from microphone array measurements in the short-time Fourier transform (STFT) domain. Let $\mathbf{y}(t,f) \in \mathbb{C}^M$ denote the microphone observations at time index $t$ and frequency $f$, and let $\mathbf{x}(t,f) \in \mathbb{C}^N$ represent the complex amplitudes of a sparse set of plane waves. The acoustic propagation model is given by the transfer matrix $\mathbf{H}(f) \in \mathbb{C}^{M \times N}$:
\begin{equation}
    \mathbf{y}(t,f) = \mathbf{H}(f)\,\mathbf{x}(t,f) + \mathbf{n}(t,f),
    \label{eq: signal model}
\end{equation}
where $\mathbf{n}(t,f)$ accounts for measurement noise.

Recovering $\mathbf{x}(t,f)$ from $\mathbf{y}(t,f)$ is an inverse problem as defined in \eqref{eq: signal model}. Since this problem is typically underdetermined, direct inversion becomes ill-posed and unstable \cite{candes2008introduction}. Sparse plane-wave decomposition aims to identify the smallest set of plane waves that sufficiently represent the observed field \cite{jin2017sound}. This can be formulated as the following constrained optimization problem:
\begin{equation}
\hat{\mathbf{x}}(t,f)=\arg\min_{\mathbf{x}(t,f)}\|\mathbf{x}(t,f)\|_{2,p}
\;\text{s.t.}\;
\mathbf{y}(t,f)=\mathbf{H}(f)\mathbf{x}(t,f),
\end{equation}
where the mixed $\ell_{2,p}$-norm promotes sparsity in the recovered coefficients \cite{jin2017sound}.

\subsection{Combined SMA–LMA Processing}
\label{subsec: combined array processing}

A straightforward strategy for exploiting hybrid arrays is to concatenate SMA and LMA observations into a single SR problem. While such concatenation has been shown to be effective in multi-SMA configurations \cite{tang2022wave}, it performs poorly for SMA–LMA hybrids. This is because LMA channels are highly sensitive to room reverberation, and when combined directly, they introduce spurious components that degrade SR performance. To mitigate this, our prior work proposed a residue refinement method, in which an initial SMA-based estimate is refined using LMA residuals \cite{Xu2025}. This two-stage design improved robustness and will be considered as a baseline in this work. Nevertheless, both concatenation and RR remain ad hoc treatments of the hybrid array geometry, motivating the unified modal framework proposed here.

\section{Proposed Method: Unified Modal Solution via SVD}
\label{sec:proposed}

\subsection{Transfer Operator and Modal Decomposition}
\label{subsec:modal-decomposition}

The reconstruction of sound fields from microphone array measurements can be framed in the operator-theoretic formulation \cite{fazi2010sound}. Let $\mathcal{D}$ denote a set of observation points in the field (e.g., plane-wave directions on a sphere), and let $\mathcal{A}$ denote the array geometry (e.g., microphone positions of a hybrid SMA–LMA array). The transfer operator
\begin{equation}
    \mathcal{H}(f): \ell^2(\mathcal{D}) \;\longrightarrow\; \ell^2(\mathcal{A})
\end{equation}
maps plane-wave coefficients defined on $\mathcal{D}$ to microphone observations at $\mathcal{A}$, where $f$ denotes frequency. Discretizing $\mathcal{D}$ into $N$ candidate directions and $\mathcal{A}$ into $M$ microphones yields the transfer matrix $\mathbf{H}(f) \in \mathbb{C}^{M\times N},$ whose entries are determined by the free-field Green’s function between direction $n$ and microphone $m$. 

\begin{figure}[t]
  \centering
  \subfigure[]{\includegraphics[width=0.4\columnwidth]{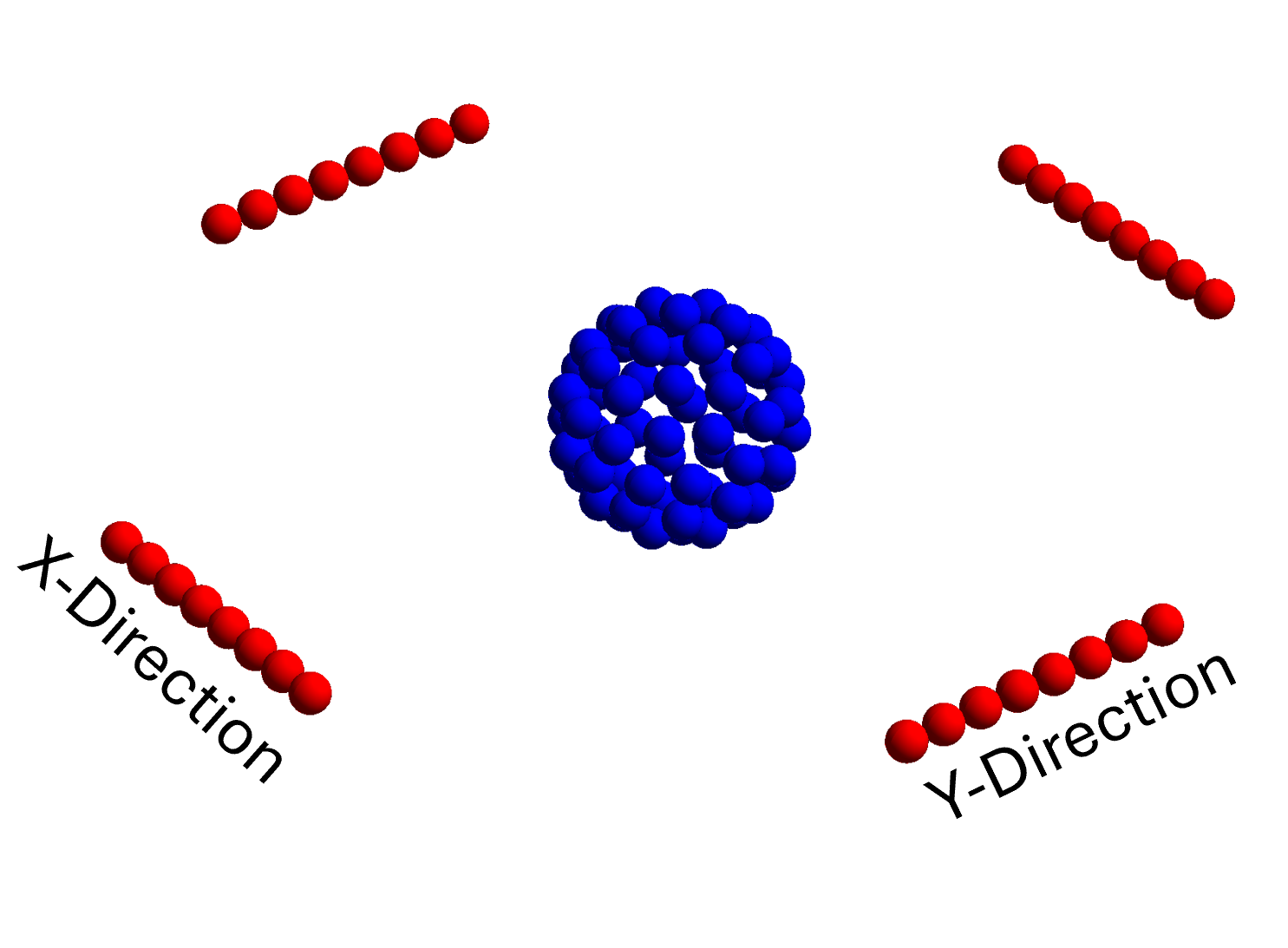}\label{fig:array setup}}
  \subfigure[]{\includegraphics[width=0.4\columnwidth]{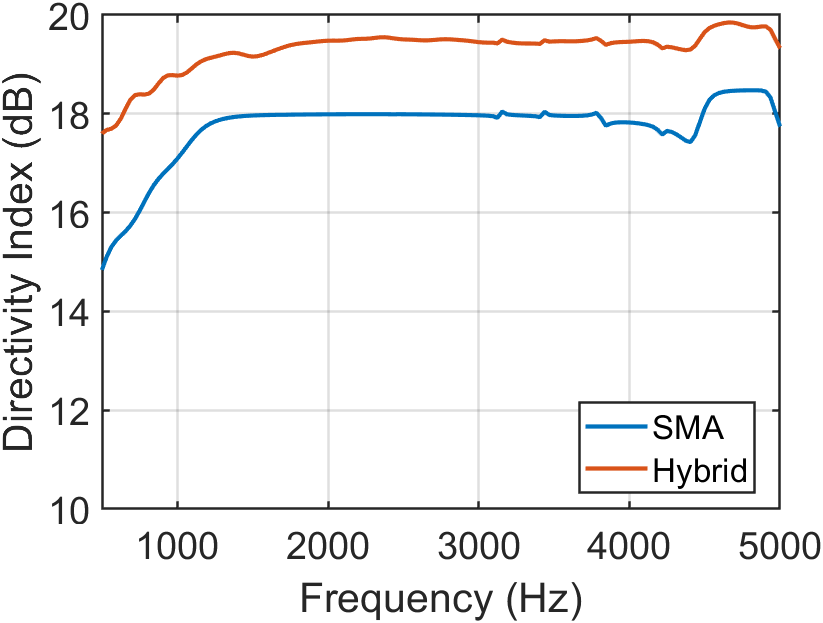}\label{fig:array di}}
  \caption{ (a) Hybrid array geometry: a 64-element open SMA of radius 10\,cm (blue) surrounded by four 8-element LMAs (red) with 4\,cm spacing, symmetrically placed 0.5\,m from the SMA center along the $x$- and $y$-axes; (b) directivity index \cite{jin2013design} comparison showing that the hybrid array achieves improved spatial selectivity over the SMA-only configuration.}
  \label{fig:array configuration}
\end{figure}

In our setup, $\mathcal{D}$ corresponds to unit-sphere directions and $\mathcal{A}$ to the hybrid array comprising a 64-element SMA and four 8-element LMAs (Fig.\ref{fig:array setup}).  More generally, $\mathcal{D}$ and $\mathcal{A}$ can be arbitrary, making the formulation agnostic to array topology and observation sampling.

Since $\mathbf{H}(f)$ is compact and ill-conditioned, direct inversion is unstable. To address this, we apply SVD to obtain an orthogonal set of modes that diagonalize the operator and order them by coupling strength. This yields a physically interpretable modal basis that links the microphone domain to the sound field domain and provides a well-conditioned dictionary for sparse recovery.
\begin{equation}
    \mathbf{H}(f) = \mathbf{U}(f)\,\mathbf{\Sigma}(f)\,\mathbf{V}^H(f),
\end{equation}
where $\mathbf{U}(f)\in \mathbb{C}^{M\times M}$ contains orthogonal microphone modes, $\mathbf{V}(f)\in \mathbb{C}^{N\times N}$ contains orthogonal field modes, and $\mathbf{\Sigma}(f)=\text{diag}(\sigma_1(f),\dots,\sigma_r(f))$ holds singular values ordered $\sigma_1(f)\geq\sigma_2(f)\geq\cdots \geq 0$. Truncating to the $K$ dominant singular values yields the reduced-rank model
\begin{equation}
    \mathbf{H}(f) \approx \mathbf{U}_K(f)\,\mathbf{\Sigma}_K(f)\,\mathbf{V}_K^H(f),
\end{equation}
where we select $K=\{9,16,25\}$ in this work, corresponding to SH orders 2–4. Projecting the observations onto $\mathbf{U}_K(f)$ and whitening by $\mathbf{\Sigma}_K(f)$ yields a stable dictionary:
\begin{equation}
    \tilde{\mathbf{y}}(t,f) = \mathbf{U}_K^H(f)\,\mathbf{y}(t,f), 
    \qquad
    \tilde{\mathbf{H}}(f) = \mathbf{\Sigma}_K^{-1}(f)\,\mathbf{V}_K^H(f).
\end{equation}
The sparse coefficients are then estimated as
\begin{equation}
    \hat{\mathbf{x}}(t,f) = \arg\min_{\mathbf{x}(t,f)} \|\mathbf{x}(t,f)\|_{2,p} 
    \; \text{s.t.}\; \tilde{\mathbf{y}}(t,f) = \tilde{\mathbf{H}}(f)\,\mathbf{x}(t,f).
\end{equation}

By construction, the modal dictionary $\tilde{\mathbf{H}}(f)$ is orthogonal and ordered by coupling strength, ensuring robust recovery even in reverberant conditions. For SMA-only setups, this reduces to
spherical-harmonic processing, while for hybrid SMA–LMA arrays it exploits the complementary spatial information of both array types.

\subsection{Modal Analysis}

\begin{figure}[t]
  \centering

  \subfigure[]{\includegraphics[width=0.44\columnwidth]{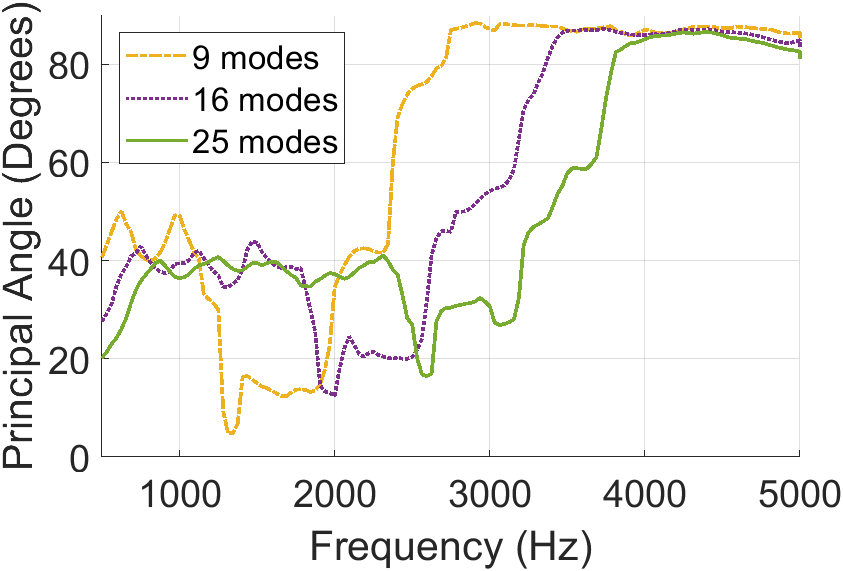}}\label{fig:subspace-analysis-a}
  \hfill
  \subfigure[]{\includegraphics[width=0.44\columnwidth, height=0.125\textheight]{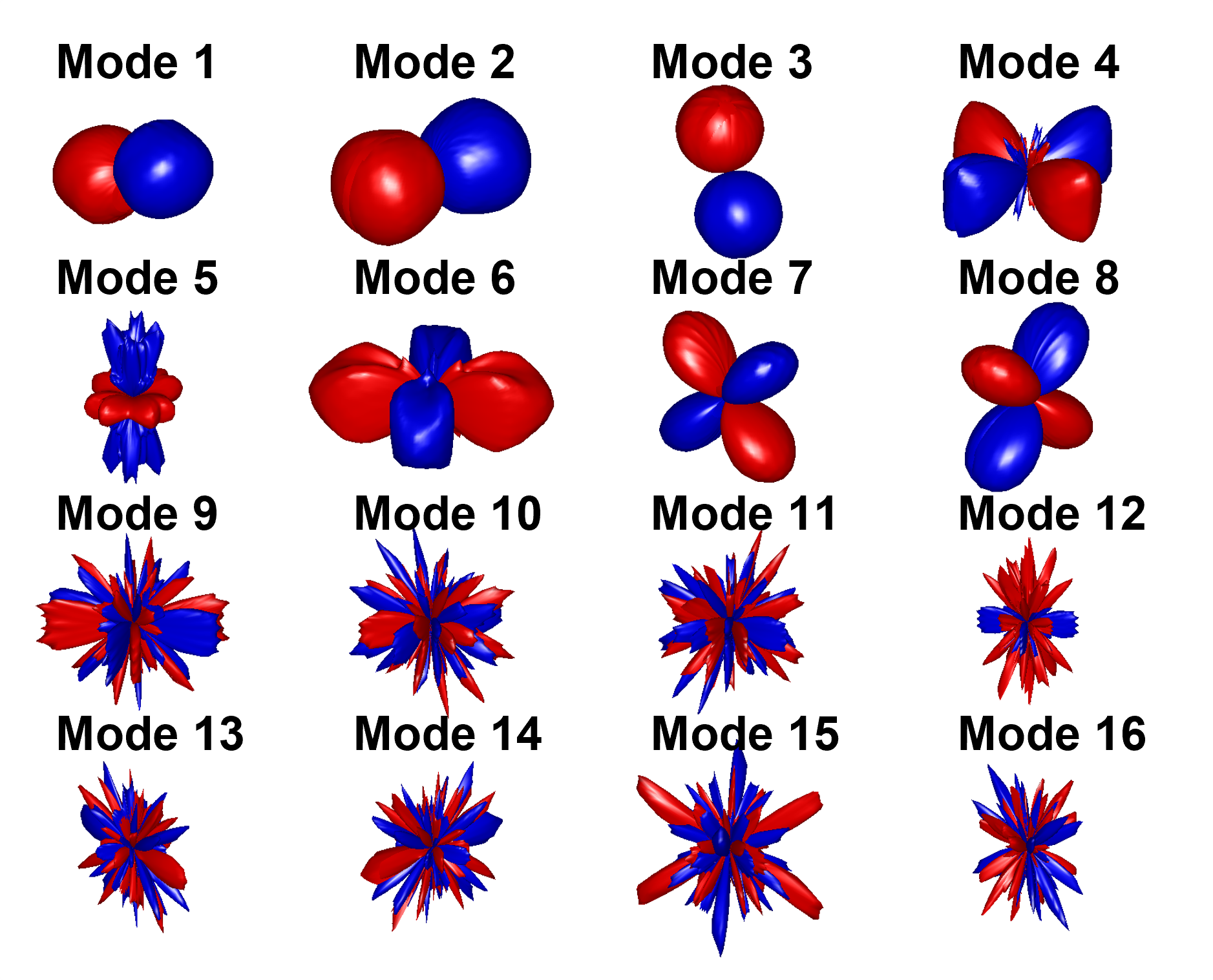}\label{fig:subspace-analysis-b}}

 \hspace{0.01\textwidth}

  \subfigure[]{\includegraphics[width=0.44\columnwidth, height=0.125\textheight]{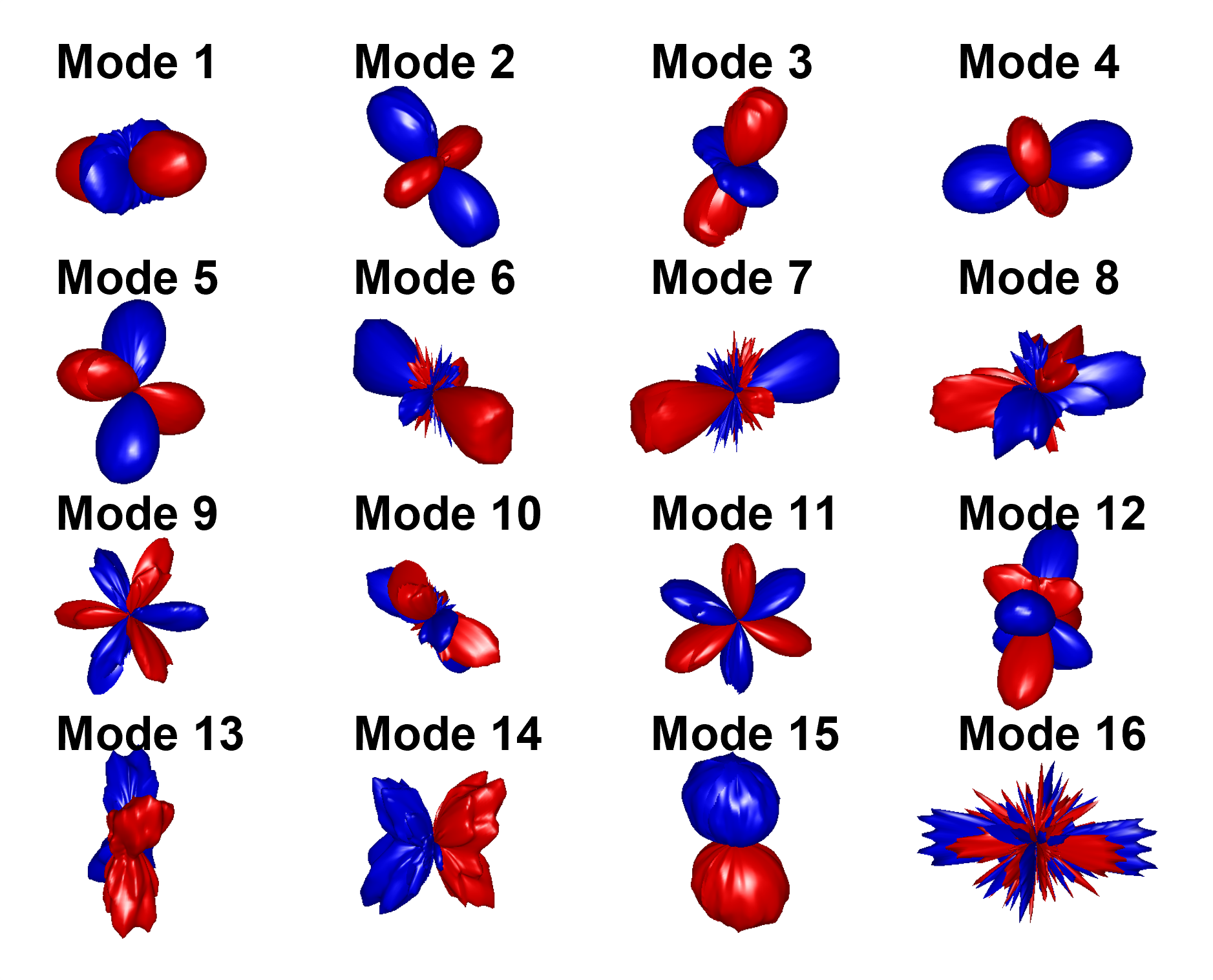}\label{}}
  \hfill
  \subfigure[]{\includegraphics[width=0.44\columnwidth, height=0.125\textheight]{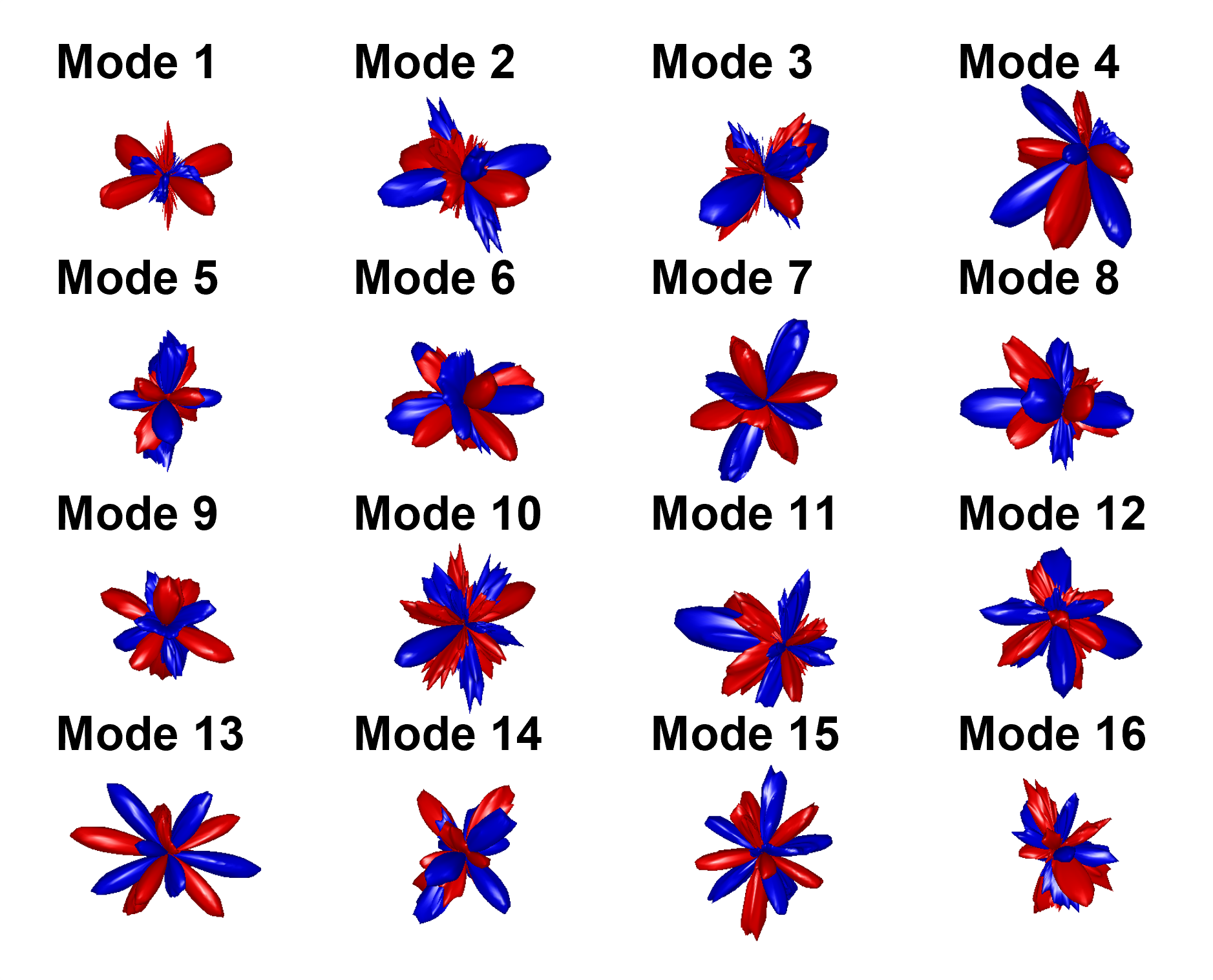}\label{fig:subspace-analysis-d}}

  \caption{Modal analysis of the hybrid SMA–LMA array. 
(a) Mean principal angle between SH and SVD subspaces, showing frequency-dependent divergence. (b)–(d) Representative SVD modes at 1.5, 2.0, and 3.5\,kHz, illustrating the frequency-dependent spatial patterns of the first 16 modes.}
  \label{fig:subspace-analysis}
\end{figure}

To investigate the properties of the derived modal basis, we analyze the subspace spanned by SVD modes relative to the classical SH basis using the principal angles \cite{knyazev2002principal}. Let $\mathcal{U},\mathcal{V}\subset \mathbb{C}^N$ be $d$-dimensional subspaces with orthonormal basis matrices $\mathbf{U},\mathbf{V}\in \mathbb{C}^{N\times d}$. The principal angles $\{\theta_i\}_{i=1}^d$ between $\mathcal{U}$ and $\mathcal{V}$ are defined through the singular values of $\mathbf{U}^H\mathbf{V}$ as
\begin{equation}
    \cos \theta_i = \sigma_i(\mathbf{U}^H\mathbf{V}), \quad i=1,\dots,d,
\end{equation}
and the mean principal angle is
\begin{equation}
    \bar{\theta} = \frac{1}{d}\sum_{i=1}^d \cos^{-1}\!\big(\sigma_i(\mathbf{U}^H\mathbf{V})\big)\cdot \frac{180}{\pi}.
\end{equation}
A larger $\bar{\theta}$ indicates greater divergence between the modal subspaces. 

Fig.\,\ref{fig:subspace-analysis}(a) shows the principal angle between SH subspaces of order $2$–$4$ (corresponding to $9$, $16$, and $25$ modes) and the SVD modes of the hybrid array. At low frequencies, SVD modes deviate strongly due to the weak excitation of spherical Bessel functions. As frequency increases, SH modes become fully excited and align more closely with SVD. Beyond the aliasing limit, SH modes suffer from spatial aliasing and fail to represent the field, whereas SVD continues to select stable modes. Figs.\ref{fig:subspace-analysis}(b)-(d) provide examples of the spatial patterns of the first 16 SVD modes at $1.5$ kHz, $2.0$ kHz, and $3.5$ kHz.

\section{Evaluation}

\subsection{Experiment Setup}
We use the hybrid array illustrated in Fig.\ref{fig:array setup}. To evaluate performance under reverberant conditions, we simulate a $10 \times 8 \times 3$\,m room using MCRoomSim \cite{wabnitz2010room} with RT60 = 0.3\,s. Plane-wave sources of 4\,s speech are generated from random directions at source–array distances of 1.5, 2.5, and 3.5\,m. The number of sources ranges from 2 to 10, with 100 trials per case. Microphone signals are obtained by convolving with the corresponding room impulse responses (RIRs) and adding spatially uncorrelated white Gaussian noise at 30\,dB SNR.  

Sparse recovery is carried out using the iterative reweighted least squares (IRLS) algorithm with dynamic regularization estimated from diffuseness \cite{epain2016spherical}. The solver is initialized with $\ell_1$-norm minimization for 10 iterations and then switched to $\ell_p$-norm ($p=0.7$) following \cite{daubechies2010iteratively}. The SR dictionary $\mathbf{H}$ is constructed from 642 uniformly sampled directions using recursive icosahedral subdivision.

\subsection{Sparse Recovery Metrics}
\label{subsec: Sparse Recovery Metrics}
\begin{figure}[t]
  \centering
  \subfigure[]{\includegraphics[width=0.45\columnwidth]{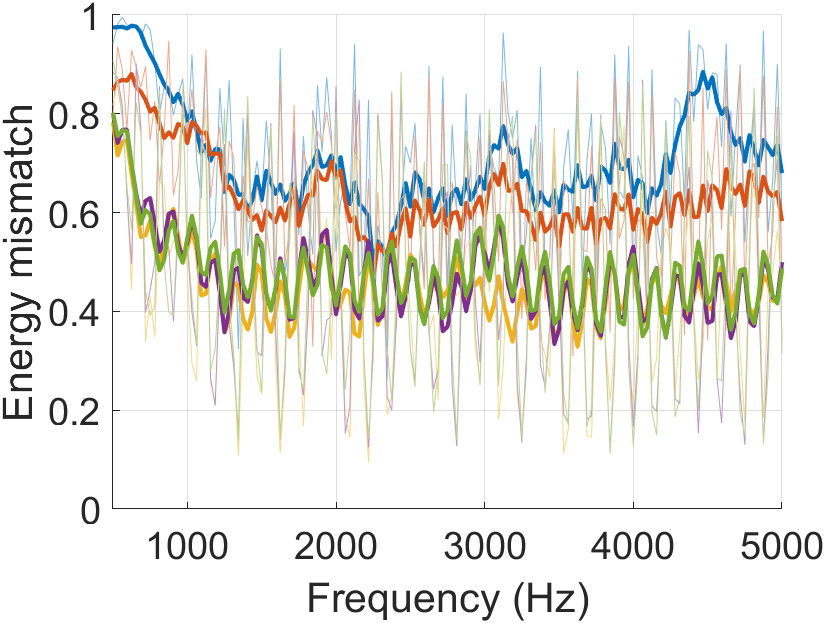}}
  \subfigure[]{\includegraphics[width=0.45\columnwidth]{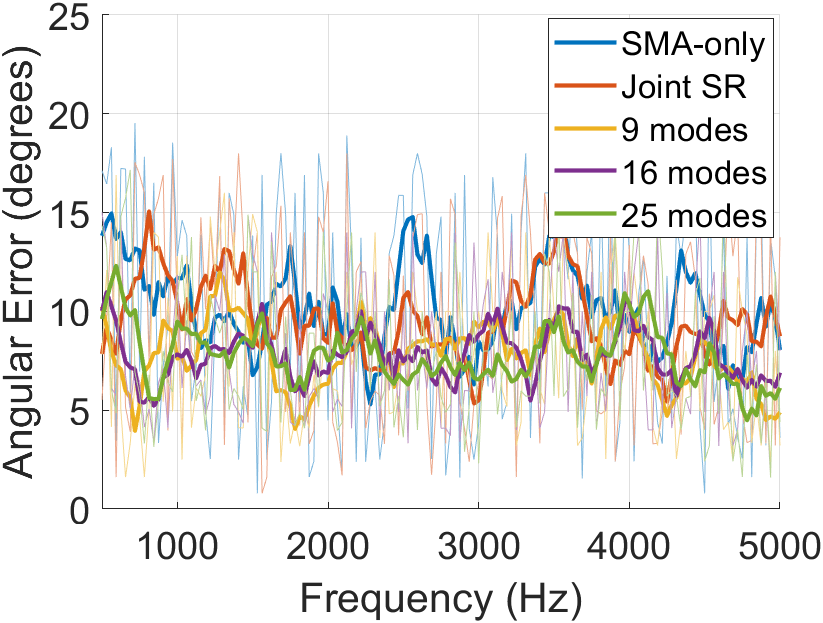}}
  \caption{Sparse recovery in a reverberant room (RT60 = 0.3 s) with 10 sources. (a) Energy map mismatch and (b) angular error versus frequency. Modal solutions consistently reduce mismatch and achieve lower angular error compared with SMA-only and joint SR.}
  \label{fig:SR vs freq}
\end{figure}

\textbf{Energy Map Mismatch:}  Following \cite{jin2017sound}, the mismatch between two energy maps is defined as
\begin{equation}
    E = \frac{K_{11}+K_{22}-2K_{12}}{K_{11}+K_{22}},
\end{equation}
with
\begin{equation}
    K_{ij} = \sum_{q=1}^{Q}\sum_{p=1}^{P}\sqrt{\rho^{(i)}_q\rho^{(j)}_p}\,
    k\!\left(\mathbf{\Omega}_q^{(i)},\mathbf{\Omega}_p^{(j)}\right),
\end{equation}
where $\rho_q^{(i)}$ and $\mathbf{\Omega}_q^{(i)}$ denote the power and direction of the $q$-th point in map $i$. The kernel function is
\begin{equation}
    k\!\left(\mathbf{\Omega}_q^{(i)},\mathbf{\Omega}_p^{(j)}\right) =
    \max\!\left(1-\frac{\angle(\mathbf{\Omega}_q^{(i)},\mathbf{\Omega}_p^{(j)})}{\pi/12},\,0\right),
\end{equation}
which linearly decays from 1 to 0 as the angular distance increases from $0$ to $\pi/12$, and vanishes beyond.

\textbf{Angular Error:}  
The angular error is the distance between the true direction $\mathbf{\Omega}_n$ and the estimated direction $\hat{\mathbf{\Omega}}_n$:
\begin{equation}
    \text{Angular Error} = \angle(\mathbf{\Omega}_n, \hat{\mathbf{\Omega}}_n).
\end{equation}
The estimate is the nearest peak within a $20^\circ$ neighbourhood of the true direction with energy above $-20$ dB and at least $80\%$ of the local maximum.

\begin{figure*}[t]
  \centering
  \subfigure[]{\includegraphics[width=0.28\textwidth]{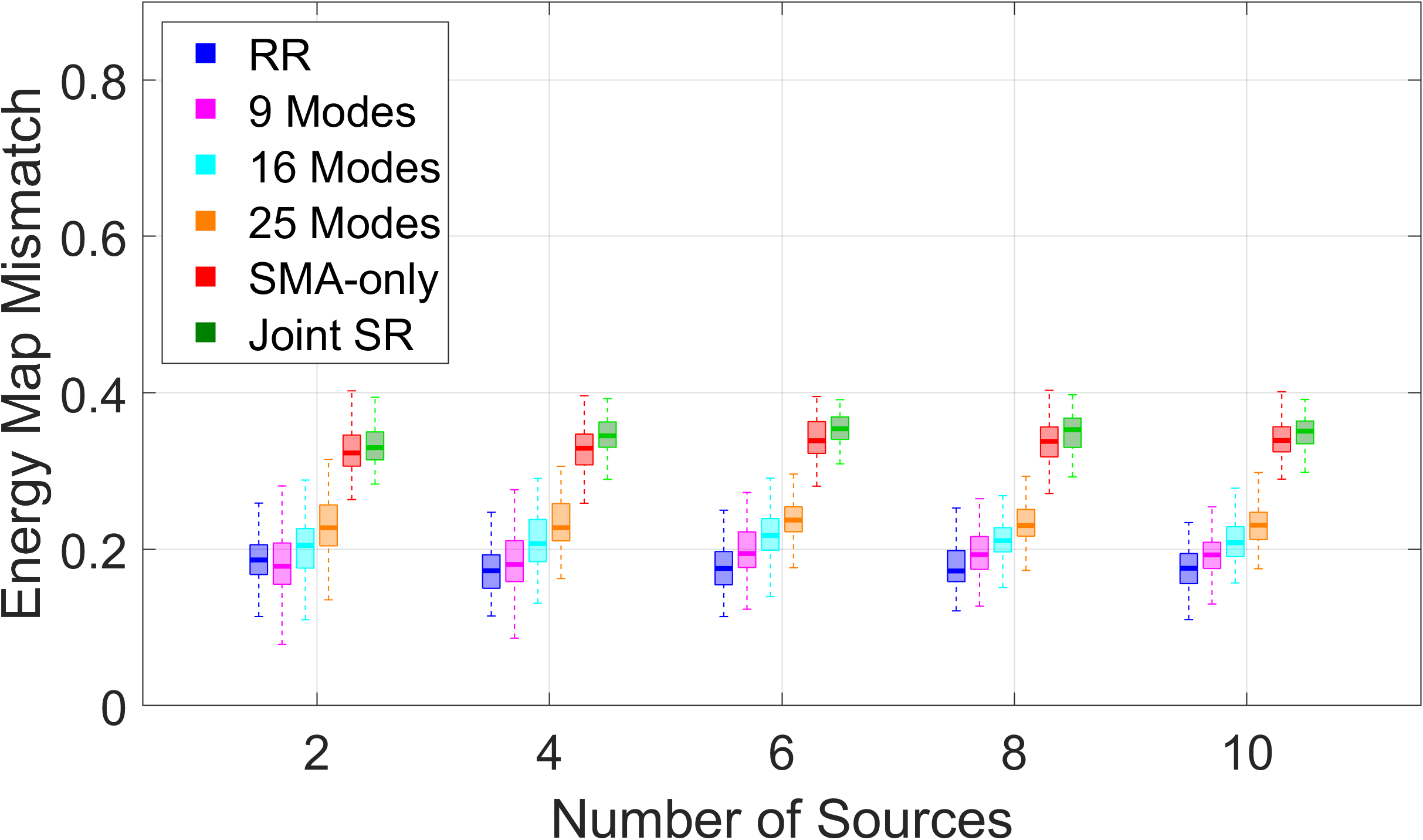}}
  \subfigure[]{\includegraphics[width=0.28\textwidth]{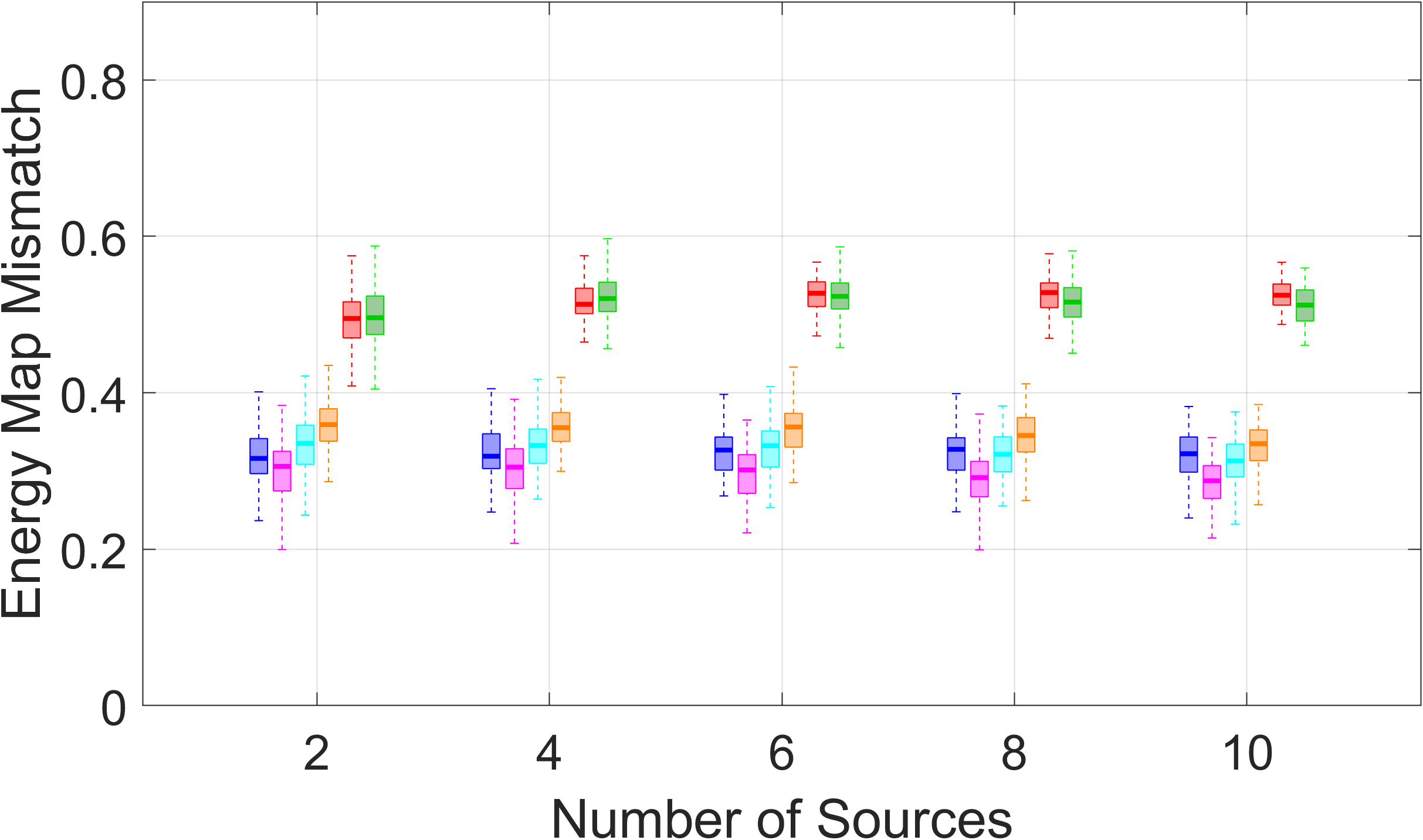}}
  \subfigure[]{\includegraphics[width=0.28\textwidth]{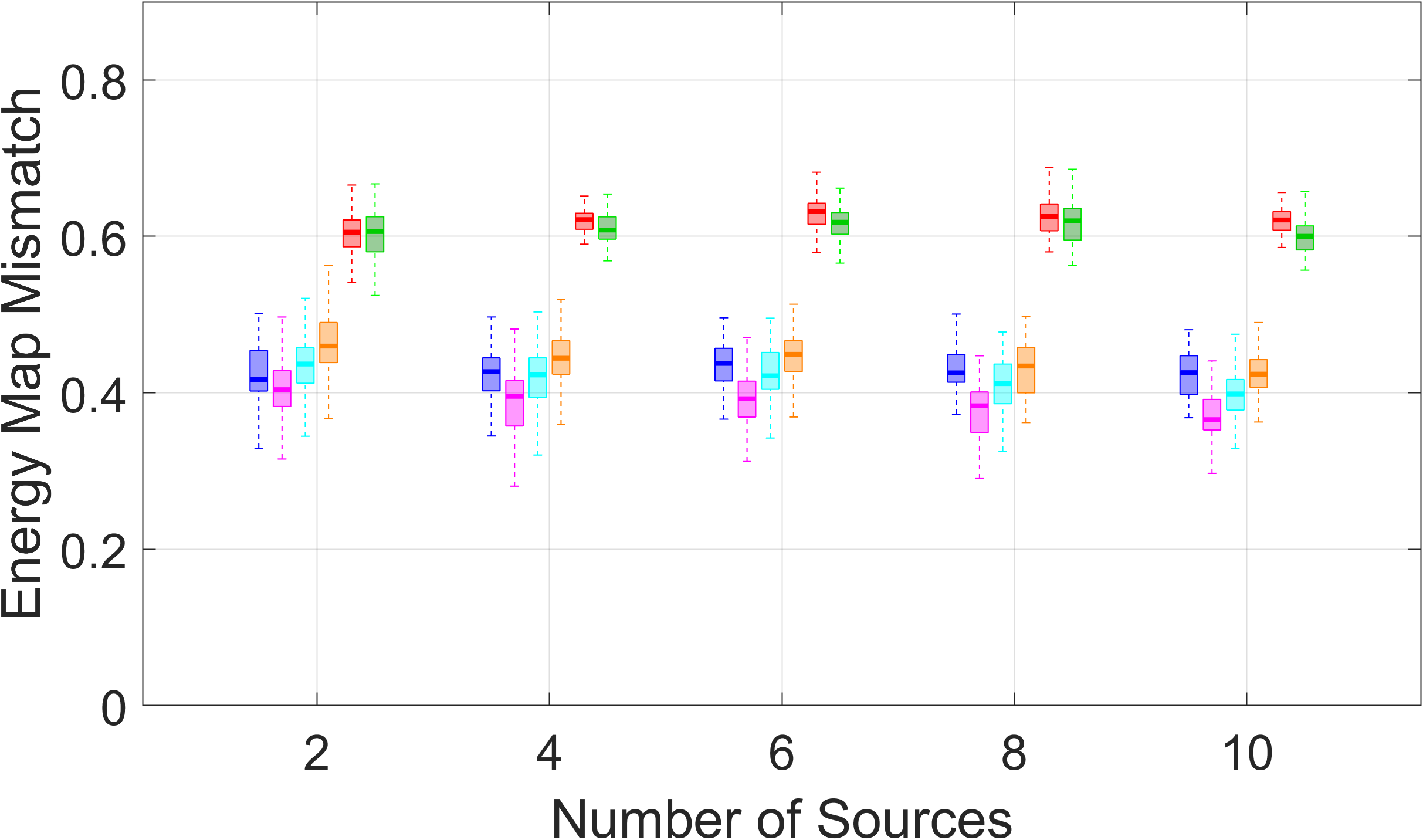}}
  \caption{Energy mismatch across different source distances (a)\,1.5m (b)\,2.5m (c)\,3.5m. Legend indicates the processing methods.}
  \label{fig:energy_mismatch room1}
\end{figure*}

\begin{figure*}[t]
  \centering
  \subfigure[]{\includegraphics[width=0.28\textwidth]{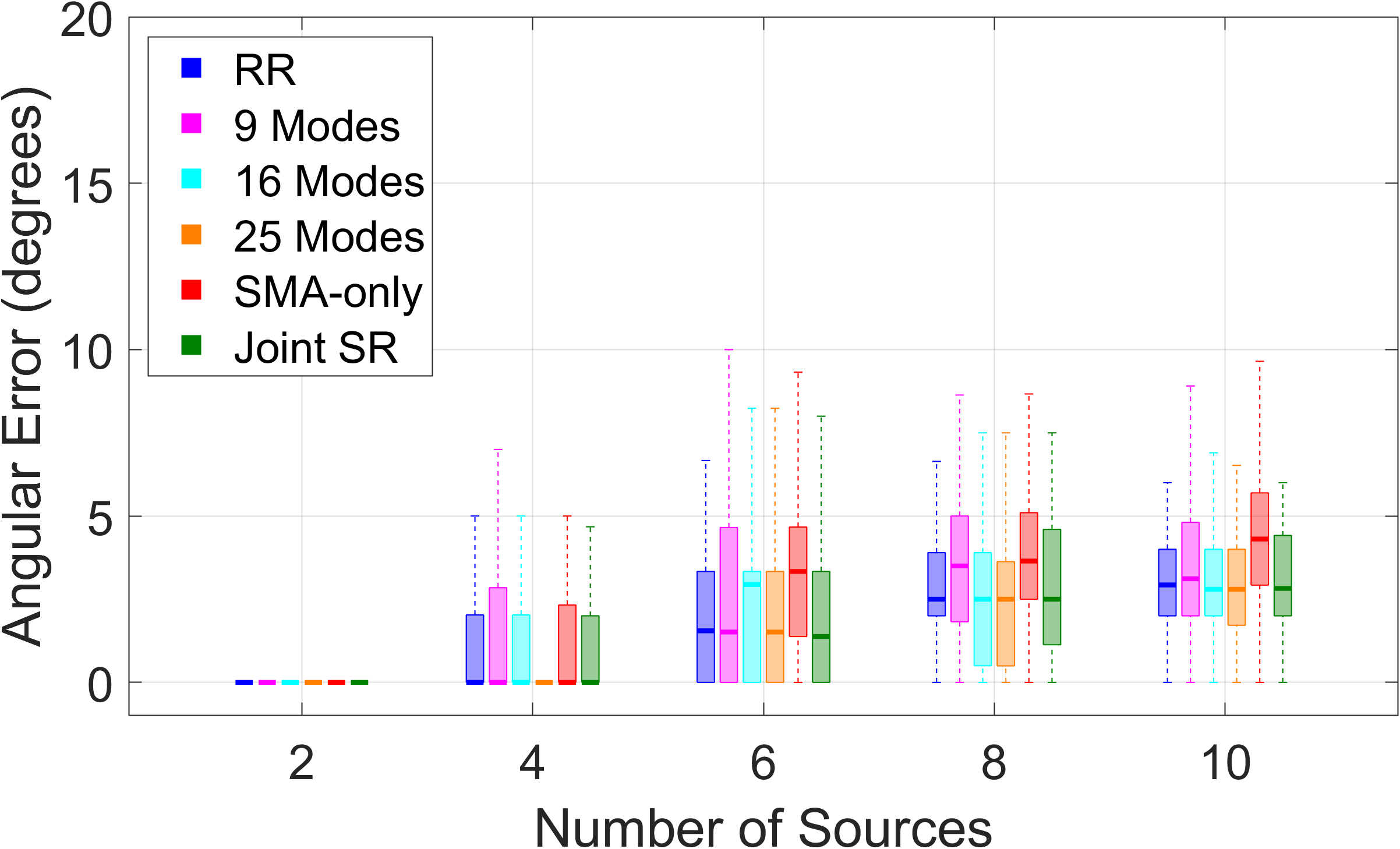}}
  \subfigure[]{\includegraphics[width=0.28\textwidth]{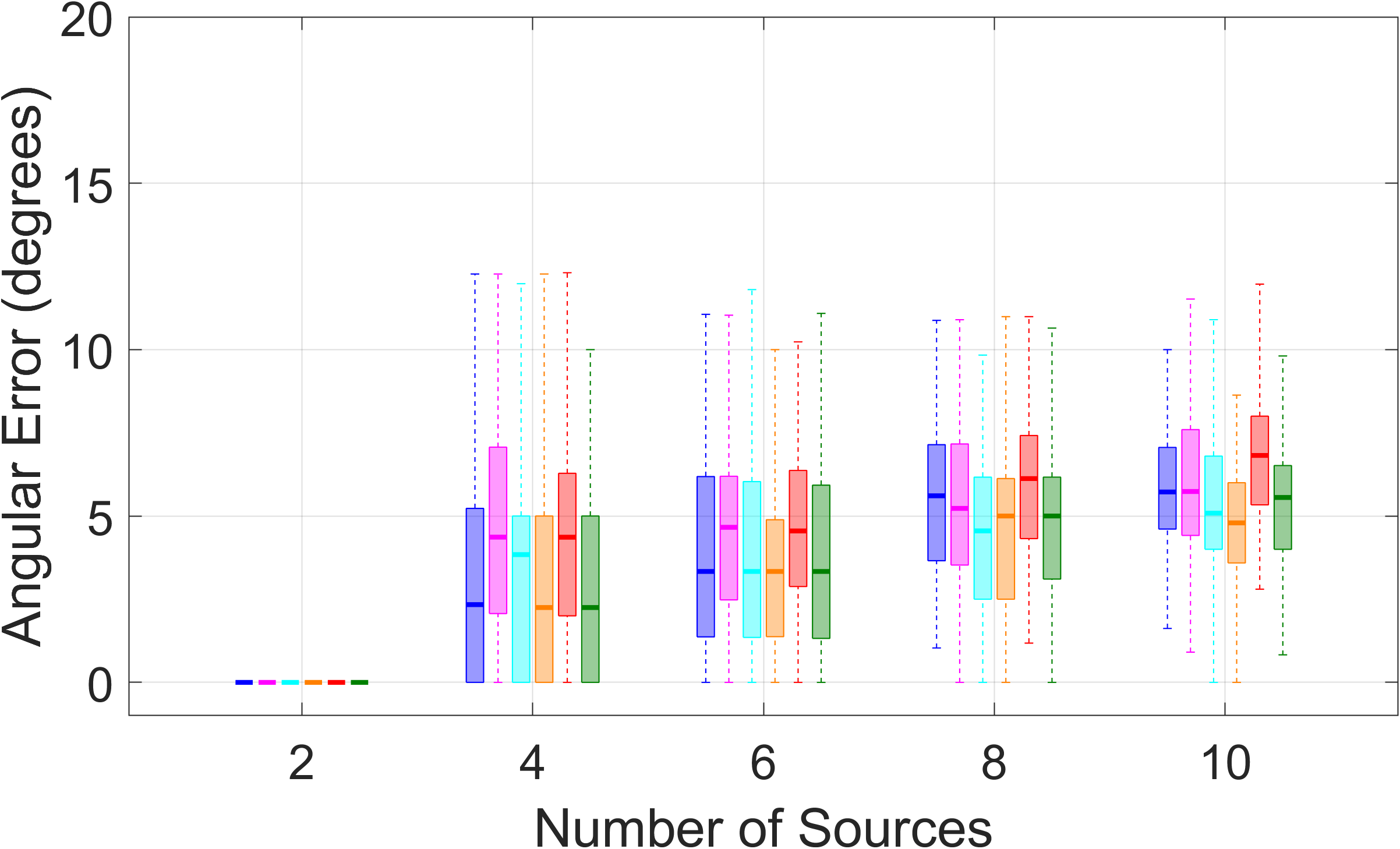}}
  \subfigure[]{\includegraphics[width=0.28\textwidth]{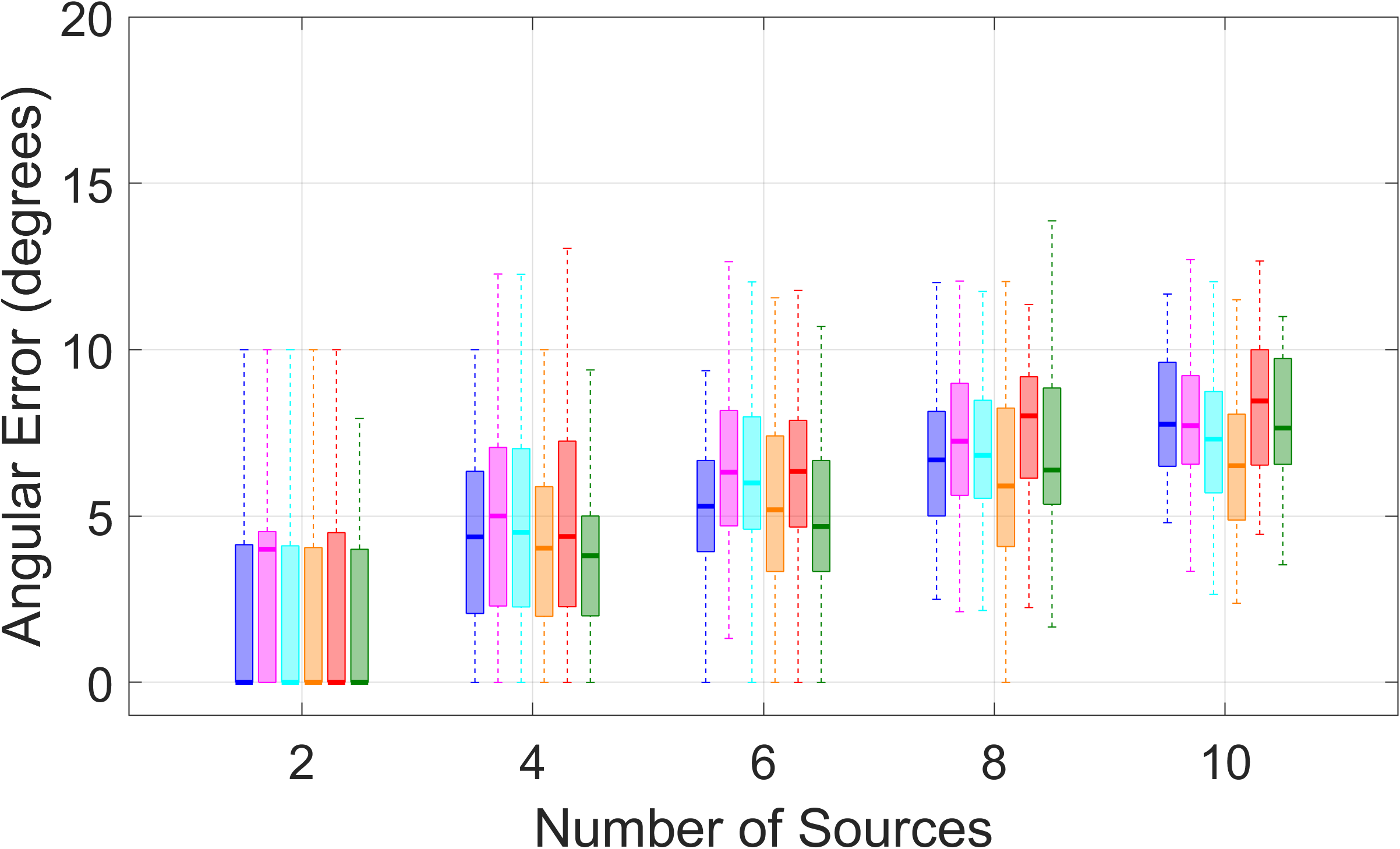}}
  \caption{Angular error across different source distances (a)\,1.5m (b)\,2.5m (c)\,3.5m. Legend indicates the processing methods.}
  \label{fig:angular_error room1}
\end{figure*}

Fig. \ref{fig:SR vs freq} compares the SR performance of SMA-only, joint SR, and the proposed modal solutions (9, 16, 25 modes) across frequency. A $200$\,Hz moving-average window is applied to smooth fluctuations for clarity. The energy map mismatch shows a clear reduction when using modal solutions, with consistently lower values across the band compared to SMA-only and joint SR, confirming improved spatial accuracy. For angular error, the differences are less visually pronounced, yet the modal solutions maintain an overall advantage, yielding smaller errors on average. These results highlight that the unified modal basis provides more stable and accurate recovery, especially in reverberant conditions.

\subsection{Robustness Evaluation}
\label{subsec: Robustness Evaluation}
We evaluate four methods: SMA-only (64-element open array), joint SMA–LMA SR, residue refinement (RR) \cite{Xu2025}, and the proposed modal solution using the first 9, 16, or 25 modes (corresponding to 2–4 order SHs). Fig.\ref{fig:energy_mismatch room1} shows energy mismatch results for three source distances ($1.5,2.5,3.5$ m). The proposed SVD-based hybrid SMA–LMA approach (pink, cyan, orange) consistently achieves lower mismatch than SMA-only (red) and joint SR (green), and is comparable to residue refinement (blue). A lower mismatch indicates higher fidelity in reproducing the spatial energy distribution. Within the proposed method, increasing the number of modes slightly raises mismatch, as weaker modes with smaller singular values are more sensitive to noise and reverberation. Nevertheless, for practical selections (9–25 modes) the framework yields substantially lower mismatch than SMA-only and joint SR, confirming robustness and accuracy under reverberant conditions.

Fig.\ref{fig:angular_error room1} shows the localization errors. The proposed method, joint SR and RR achieve comparable accuracy, consistently outperforming SMA-only across all conditions, due to the inclusion of LMA data. Performance degrades with source distance, reflecting the reduced resolution of far-field arrays. For the proposed method, increasing the number of modes yields a further reduction in angular error, indicating that additional modes capture finer spatial detail that improves localization accuracy. Taken together, these results highlight a trade-off: fewer modes favor energy-map fidelity, while more modes enhance localization accuracy, suggesting future work on task-dependent optimal mode selection.

\section{Conclusion}
We presented a data-driven SVD-modal framework for sparse recovery with hybrid SMA–LMA arrays. The method generalizes SH (reducing to SH for SMA-only) and, with LMAs, introduces complementary, well-conditioned modes. Modal analysis reveals the frequency-dependent divergence from SH and confirm the resulting gains in spatial resolution. In reverberant tests, the proposed approach consistently outperforms SMA-only and direct concatenation, and achieves performance on par with the heuristic residue-refinement baseline, while providing a principled, unified framework that generalizes beyond ad hoc designs. Future work will focus on algorithms for task-dependent, optimal mode selection.

\newpage
\bibliographystyle{IEEEtran}
\bibliography{strings,references}

\end{document}